\documentclass[a4paper]{article}
\usepackage{xcolor}
\usepackage{multirow}
\usepackage{amsmath}
\usepackage{graphicx}
\usepackage{setspace}
\usepackage{mdwlist}
\usepackage{enumitem}
\usepackage{booktabs}
\usepackage{geometry}
\usepackage{epstopdf}
\usepackage{cleveref}
\usepackage{afterpage} 
\usepackage{float}
\usepackage{subcaption}
\usepackage{rotating}
\usepackage{adjustbox}

\usepackage{INTERSPEECH2022}

\title{Learning from human perception to improve automatic speaker verification in style-mismatched conditions}
\name{Amber Afshan\thanks{This work was supported in part by the NSF.}, Abeer Alwan}
\address{
  Department of Electrical and Computer Engineering, University of California Los Angeles, USA}
\email{amberafshan@g.ucla.edu, alwan@g.ucla.edu}

\begin{document}

\maketitle
\begin{abstract}
Our prior experiments show that humans and machines seem to employ different approaches to speaker discrimination, especially in the presence of speaking style variability. The experiments examined read versus conversational speech. Listeners focused on speaker-specific idiosyncrasies while ``telling speakers together'', and on relative distances in a shared acoustic space when ``telling speakers apart''. However, automatic speaker verification (ASV) systems use the same loss function irrespective of target or non-target trials. To improve  ASV performance in the presence of style variability, insights learnt from human perception are used to design a new training loss function that we refer to as ``$C_\textrm{llr}$CE loss''. $C_\textrm{llr}$CE loss uses both speaker-specific idiosyncrasies and relative acoustic distances between speakers to train the ASV system. When using the UCLA speaker variability database, in the x-vector and conditioning setups, $C_\textrm{llr}$CE loss results in significant relative improvements in EER by 1-66\%, and minDCF by 1-31\% and 1-56\%, respectively, when compared to the x-vector baseline. Using the SITW evaluation tasks, which involve different conversational speech tasks, the proposed loss combined with self-attention conditioning results in significant relative improvements in EER by 2-5\% and minDCF by 6-12\% over baseline. In the SITW case, performance improvements were consistent only with conditioning. 
\end{abstract}


\section{Introduction \label{sec:introduction}}
Automatic speaker verification (ASV) is an open-set problem, i.e., test speakers are unavailable to the system during training but available during enrollment. ASV is, hence, a metric learning problem that maps speakers to a discriminative embedding space. Most of the work on speaker verification has focused on training with identification objectives. One such identification objective is cross-entropy loss~\cite{krizhevsky_imagenet_2012, Goodfellowetal2016}. Identification loss functions learn linearly separable embeddings by focusing on maximizing inter-speaker distances. However, they do not typically minimize intra-speaker distances. Hence, the resulting embeddings do not have adequate discriminative properties. 

To address the drawbacks of identification loss in ASV systems, Angular softmax~\cite{li_angular_2018} loss was used. Angular softmax uses cosine similarity as the logit input to the softmax layer. Additive margin variants of Angular softmax such as AM-Softmax~\cite{wang_cosface_2018, wang_additive_2018} and AAM-Softmax~\cite{Deng_2019_CVPR} use a cosine margin penalty on the target logit. These techniques although effective,  have been proven sensitive to the value of scale and margin. 

As an alternative to identification objectives, metric learning objectives that focus on minimizing intra-speaker distances have been used. Metric learning objectives such as contrastive loss~\cite{chen_compensation_2012} and triplet loss~\cite{schroff_facenet_2015} have been used in ASV tasks with some success~\cite{zhang_text-independent_2018, chowdhury_attention-based_2018}. However, these approaches require careful selection of triplet pairs i.e. anchor, positive and negative pairs, resulting in longer training cycles. Apart from the high computational cost, these losses do not consider the performance measures (such as equal error rate (EER) and detection cost function (DCF)) in training; these measures are used in the final evaluation of the speaker verification task. 

It has been shown that considering a metric related to the final evaluation improves ASV performance further at least in text dependent ASV systems by using aAUC~\cite{mingote_optimization_2020}, aDCF~\cite{mingote_optimization_2019} and $C_\textrm{llr}$~\cite{mingote_log-likelihood-ratio_2021} objectives. The $C_\textrm{llr}$ loss, in particular, provides performance improvements without the need for triplet pairs and provides computational cost similar to that of identification objectives such as cross-entropy loss. $C_\textrm{llr}$  was evaluated in a text dependent speaker verification task~\cite{mingote_log-likelihood-ratio_2021} and its efficacy has not been evaluated in a text independent case. 

Given that everyday style variations in speech affect both inter- and intra-speaker variabilities~\cite{shriberg_does_2009, chen_compensation_2012}, it is important to use a loss function that maximizes inter-speaker distances and minimizes intra-speaker distances. To addresses this issue, in this paper, we introduce a loss function that is inspired by human speech perception.


\subsection{Comparison between Humans and Machines}
Speaking style variations occur frequently in everyday situations such as having a conversation, giving instructions, talking to a pet, etc. However, these variations have little effect on human ability to recognize a familiar voice~\cite{wenndt_machine_2012}. Previously it has been shown that familiarity has an influence on human strategy to recognize talkers: familiar talkers are recognized by matching the stimuli to stored voice templates, while unfamiliar talkers are recognized through acoustic feature comparisons~\cite{van_lancker_voice_1987}. 

Humans have shown to outperform machines in a task of discriminating unfamiliar speakers in both style-matched and -mismatched conditions from samples of read and pet-directed speech (characterized by exaggerated prosody)~\cite{parktowards2018, parktarget2019}. In our recent experiments~\cite{afshan_speaker_2020}, results suggest that humans and machines maybe employing different approaches to speaker discrimination in cases of moderate style variability. Moreover, two studies~\cite{lavanhow2019, johnsoncomparing2020} have shown that humans vary their perceptual strategies when ``telling people together'' versus ``telling people apart.'' On the other hand, machines apply the same approach irrespective of target or non-target trials~\cite{parktarget2019}. Given that humans and machines seem to employ different approaches to speaker discrimination, it is possible that machines might do better if they employed human perceptual strategies. In addition, humans might do better with machine assistance in certain situations. Therefore, we focus on learning from human speaker perceptual strategies in developing ASV algorithms, in particular, introducing a new training loss function.

In this work, we propose the $C_\textrm{llr}$CE loss function for text-independent ASV, especially in cases of style-mismatch. This loss function is inspired by strategies used by humans for an unfamiliar speaker discrimination task in the presence of moderate style variability (read versus conversational speech). Section~\ref{sec:method} presents the proposed method. The experimental setup is described in Section~\ref{sec:cllr_experimental_setup}, and the results and discussion are presented in Section~\ref{sec:results}. We conclude with Section~\ref{sec:conclusion}.

\begin{table}[b]
    \centering
        \caption{UCLA SVD database statistics in terms of number of utterances. }
    \label{tab:stats}
\resizebox{\linewidth}{!}{%
\begin{tabular}{c|ccccc}
\toprule
\textbf{Style} & \textbf{read} & \textbf{instructions} & \textbf{narrative} & \textbf{conversation} & \textbf{pet-directed} \\ \midrule
\midrule
Enroll  & 200 & 204 & 625$^+$ & 197  & 35$^+$ \\ 
Test        & 199 & 204 & 625$^+$ & 174  & 35$^+$ \\ \bottomrule
\multicolumn{6}{p{\linewidth}} {$^+$ Same enroll and test utterances. }\\
\end{tabular}%
}
\end{table}

\begin{table*}[t]
\caption{Performance using the UCLA database (in EER and minDCF) with CE and $C_\textrm{llr}$CE loss functions. The loss functions are used to train the x-vector system and the best performing VFR conditioning: concatenation with gating. The best performance in each condition with a statistically significant improvement over the baseline is boldfaced. If denoted by a `*' it is not a statistically significant improvement over the baseline.}
\label{tab:cllr_loss}
\centering
\resizebox{0.9\linewidth}{!}{%
\begin{tabular}{c|c|rr|rr|rr|rr}
\toprule
\toprule
     \multicolumn{2}{c|} {\textbf{Loss}}         & \multicolumn{4}{c|}{\textbf{CE}} & \multicolumn{4}{c}{\textbf{$C_\textrm{llr}$CE}} \\
             \midrule
\multirow{2}{*} {\textbf{Enroll}}      &  \multirow{2}{*}{\textbf{Test}}         & \multicolumn{2}{c|} {\textbf{x-vector (Baseline)} }     & \multicolumn{2}{c|} {\textbf{VFR conditioning  }}    & \multicolumn{2}{c|} {\textbf{x-vector}}     & \multicolumn{2}{c} {\textbf{VFR conditioning  }}    \\
\cmidrule{3-10}
& & \textbf{EER \%}     &  \textbf{minDCF$_{0.01}$}  & \textbf{EER \% }    &  \textbf{minDCF$_{0.01}$}  & \textbf{\textbf{EER \%}}     &  \textbf{minDCF$_{0.01}$}  & \textbf{EER \%  }   &  \textbf{minDCF$_{0.01}$} \\
\midrule
\midrule
\multirow{5}{*} {read}  
&read         & 0.50  & 0.018 & 0.50  & 0.013* & 0.50  & 0.023 & 0.50  & 0.018 \\
&instructions & 0.49  & 0.054 & 0.49  & 0.037* & 0.49  & 0.037* & 0.49  & 0.027* \\
&conversation         & 2.86  & 0.254 & 2.29  & 0.232 & 2.29  & 0.240 & \textbf{1.71}  & \textbf{ 0.197} \\
&narrative      & 0.80  & 0.162 & 0.80  & 0.115* & 0.80  & 0.123* & 0.80  & 0.104* \\
&pet-directed          & 17.14 & 0.928 & \textbf{14.29} & 0.943 & 17.14 & 0.943 & 17.14 & 0.886* \\
\midrule
\multirow{5}{*} {instructions}  
&read         & 1.47  & 0.154 & \textbf{0.98}  & 0.120 & 1.47  & 0.137 & 1.47  & 0.108* \\
&instructions & 0.45  & 0.005 & 0.45  & 0.005 & 0.45  & 0.005 & 0.45  & 0.005 \\
&conversation         & 2.79  & 0.296 & 2.79  & 0.263* & 2.79  & 0.263* & \textbf{2.24}  & \textbf{0.238} \\
&narrative      & 1.23  & 0.110 & \textbf{0.77}  & 0.102 & 0.92  & 0.090 & \textbf{0.77}  & \textbf{0.072} \\
&pet-directed          & 18.92 & 0.933 & \textbf{13.51} & 0.933 & 16.22 & 0.920 & 16.22 & \textbf{0.908} \\
\midrule
\multirow{5}{*} {conversation}  
&read         & 2.03  & 0.246 & \textbf{1.52}  & 0.178 & \textbf{1.52}  & 0.188 & \textbf{1.52 } & \textbf{0.173} \\
&instructions & 2.97  & 0.267 & 2.48*  & 0.248* & 2.48*  & 0.225* & 2.48*  & 0.213* \\
&conversation         & 0.57  & 0.035 & 0.57  & 0.035 & 0.57  & 0.029* & 0.57  & 0.020* \\
&narrative      & 1.94  & 0.224 & 1.94  & 0.187 & 1.94  & 0.179 & 2.10  & \textbf{0.155} \\
&pet-directed          & 20.00 & 0.887 & \textbf{17.14 }& 0.915 & \textbf{17.14} & 0.900 & \textbf{17.14} & \textbf{0.858} \\
\midrule
\multirow{4}{*} {narrative}  
&read         & 0.48  & 0.046 & 0.32*  & 0.032* & \textbf{0.16 } & 0.036 & \textbf{0.16}  & \textbf{0.020} \\
&instructions & 0.46  & 0.024 & 0.46  & 0.019* & 0.46  & 0.019* & 0.46  & 0.013* \\
&conversation         & 1.46  & 0.132 & 1.10  & 0.121 & 1.10  & 0.127 & \textbf{0.73}  &\textbf{ 0.096} \\
&pet-directed          & 18.58 & 0.828 & \textbf{13.27} & 0.908 & 13.27 & 0.855 & 14.16 & \textbf{0.841} \\
\midrule
\multirow{4}{*} {pet-directed}  
&read         & 14.29 & 0.886 & 14.29 & 0.829* & 14.29 & 0.857* & 14.29 & 0.871* \\
&instructions & 18.92 & 0.919 & \textbf{13.51} & 0.946 & 16.22 & 0.934 & 16.22 & \textbf{0.908}\\
&conversation         & 21.21 & 0.914 & \textbf{18.18} & 0.867 & \textbf{18.18} & 0.842 & 21.21 & 0.774* \\
&narrative      & 19.47 & 0.886 & \textbf{14.16} & 0.929 & 15.93 & 0.892 & 15.04 &\textbf{ 0.864} \\
\bottomrule
\bottomrule
\end{tabular}%
}
\end{table*}

\begin{table*}[t]
\caption{Performance using the SITW evaluation (in EER and minDCF) with CE, $C_\textrm{llr}$, and $C_\textrm{llr}$CE loss functions. The loss functions are used to train the x-vector system and the best performing VFR conditioning: concatenation with gating. The best performance in each condition with a statistically significant improvement over the baseline is boldfaced.} 
\label{tab:sitw_cllr}
\centering
\resizebox{\textwidth}{!}{%
\begin{tabular}{c|c|rr|rr|rr|rr}
\toprule
\toprule
&      & \multicolumn{2}{c|}{\textbf{Core-Core}} & \multicolumn{2}{c|}{\textbf{Core-Multi}} & \multicolumn{2}{c|}{\textbf{Assist-Core}} & \multicolumn{2}{c}{\textbf{Assist-Multi}} \\
\cmidrule{3-10}
\textbf{Loss} &    \textbf{Model}               & \textbf{EER \%}     &  \textbf{minDCF$_{0.01}$}  & \textbf{EER \% }    &  \textbf{minDCF$_{0.01}$}  & \textbf{\textbf{EER \%}}     &  \textbf{minDCF$_{0.01}$}  & \textbf{EER \%  }   &  \textbf{minDCF$_{0.01}$} \\

\midrule
\midrule

\multirow{2}{*} {CE} & x-vector (Baseline)        & 3.66 & 0.3820       & 5.87 & 0.4629       & 5.47 & 0.4041       & 6.90 & 0.4512       \\
& VFR conditioning         & 3.69 & 0.3989       & 5.81 & 0.4740       & \textbf{5.26} & 0.4027       & \textbf{6.54} & 0.4651       \\

\midrule

\multirow{2}{*} {$C_\textrm{llr}$} & x-vector   & 4.13  & 0.4153       & 6.46 & 0.4940       & 6.24 & 0.4376       & 7.57 & 0.4824       \\
& VFR conditioning                  & 4.29  & 0.4009       & 6.65 & 0.4821       & 6.28 & 0.4337       & 7.68 & 0.4776       \\

\midrule

\multirow{2}{*} {$C_\textrm{llr}$CE} &  x-vector             & 3.77  & 0.3654       & 5.88 & 0.4394       & 5.70 & 0.3833       & 6.74 & 0.4290        \\
& VFR conditioning        & \textbf{3.47}  & \textbf{0.3346 }      & \textbf{5.73} & \textbf{0.4178}       & 5.36 & \textbf{0.3738} & 6.73 & \textbf{0.4191}       \\
\bottomrule
\bottomrule
\end{tabular}%
}
\end{table*}

\section{Proposed Method\label{sec:method}}

\subsection{Human speaker perception}
Our previous work~\cite{afshanspeaker2022} studied human speaker perception for moderate style variability (read versus conversational speech). The results showed that listeners find it easier to ``tell speakers together'' using speaker-specific idiosyncrasies, while listeners ``tell speakers apart'' based on relative positions within a shared acoustic structure rather than speaker-specific features. 

This work aims to incorporate this strategy in the training loss function. Thus, we need a loss function that focuses on speaker-specific idiosyncrasies for the ``target speaker'' task while using acoustic distances between speakers for the ``non-target speaker'' task. 

\subsection{Embedding Extractors}
An x-vector/PLDA system~\cite{snyder_x-vectors_2018} is the baseline used in this paper. The inputs to the embedding extractor are 30-dimensional mel-frequency cepstral coefficients (MFCCs) using a 25~ms frame length and a 10~ms frame shift. The MFCCs are mean normalized over a sliding window of up to 3~secs. Extrinsic data augmentation of noise and reverberation~\cite{snyder_x-vectors_2018} was applied to the training data.

Since, the x-vector system performance is degraded in the case of  style-mismatch~\cite{afshan_variable_2020}, we also want to evaluate the proposed method in a system that has lesser degradation due to style-mismatch. Hence, we perform additional experiments using an entropy-based variable frame rate (VFR) conditioning network~\cite{afshan2022attention, afshanspeaking2022} developed to compensate for speaking style effects. This method uses VFR output~\cite{zhu_use_2000, afshan_variable_2020, ravi_voice_2019} as a conditioning vector in the self-attention pooling layer. Five different approaches were used for conditioning. Among those, the best performing  VFR conditioning network,  concatenation with gating, is used. In this setup, the statistical pooling layer is replaced with a self-attention layer. The self-attention layer is then conditioned using an entropy-based variable frame rate vector~\cite{afshan_variable_2020}. 

\subsection{Loss Functions}

\subsubsection{Cross-Entropy (CE) Loss}
A widely-used loss function for training ASV systems, including the x-vector system, is the cross-entropy loss. This function calculates loss for a multi-class classification problem. CE loss can be calculated as,
\begin{equation}
    L_{CE} = -\frac{1}{m} \sum_{i=0}^m \log{\frac{e^{(\mathbf{W}^T_{\mathbf{y}_i}.\mathbf{x}_i+\mathbf{b}_{\mathbf{y}_i})}}{\sum_{j=0}^N e^{(\mathbf{W}^T_{j}.\mathbf{x}_j+\mathbf{b}_{j})}}} 
\end{equation}
\noindent where $\mathbf{x}_i$ is the $i^\textrm{th}$ training sample, $\mathbf{y}_i$ is the ground truth speaker label of the $i^\textrm{th}$ training sample, $i \in \{1,\dots,m\}$, where $m$ is the total number of training samples. $\mathbf{W}$ indicates the weight matrix, $\mathbf{b}$ is the bias vector. $\mathbf{W}_j$ and $\mathbf{W}_{\mathbf{y}_i}$ are the $j^\textrm{th}$ and $\mathbf{y}_i^\textrm{th}$ columns of $\mathbf{W}$, respectively. $\mathbf{b}_j$ and $\mathbf{b}_{\mathbf{y}_i}$ are the $j^\textrm{th}$ and $\mathbf{y}_i^\textrm{th}$ bias values, respectively. The CE loss is calculated for a total of $N$ speakers. 

The CE loss aims at maximizing inter-speaker distances but it does not minimize intra-speaker distances. By maximizing inter-speaker distances (the posterior probability of the correct class), the extracted embeddings are linearly separable. On the other hand, for the embeddings to include desirable discriminative features, the loss should also minimize intra-speaker distances (that is increase embedding similarity). The embeddings trained on CE loss--maximizing inter-speaker distances--are equivalent to the human approach of focusing on relative positions within a shared acoustic structure to ``tell speakers apart''. To minimize intra-speaker distances and implement other aspects of human perception strategies, we need a loss that focuses on speaker-specific idiosyncrasies.

\subsubsection{$C_\textrm{llr}$ Loss}
To focus on speaker-specific idiosyncrasies without increasing the length of the training cycles, we chose the log-likelihood-ratio cost function ($C_\textrm{llr}$)~\cite{vanleeuwenintroduction2007} as a loss function for training the embedding extractor that we refer to as ``$C_\textrm{llr}$ loss''. 

$C_\textrm{llr}$ is an application independent measure for evaluating soft decisions in ASV performance. There is a closed-form solution for $C_\textrm{llr}$~\cite{vanleeuwenintroduction2007} that provides the $C_\textrm{llr}$ loss function as follows:
\begin{align}
C_\textrm{llr}(\theta) &= \frac{1}{2}\left( \frac{C_{tar}(\theta)}{N_\textrm{tar}} +  \frac{C_{non}(\theta)}{N_\textrm{non}} \right) \\
    C_{tar}(\theta) &= \sum_{i\in \textrm{tar}} \log_2(1+e^{-s_{\theta}(\mathbf{x}_i,\mathbf{y}_i)}) \\
        C_{non}(\theta) &= \sum_{i\in \textrm{non}} \log_2(1+e^{s_{\theta}(\mathbf{x}_i,\mathbf{y}_i)}) 
\end{align}

\noindent where $\theta$ represents the model parameters, $s_{\theta}(\mathbf{x}_i,\mathbf{y}_i)$ is the score from the last layer of the embedding extractor for speaker $\mathbf{y}_i$ from input $\mathbf{x}_i$, `tar' is a set of target speakers and `non' is a set of non-target speakers. The two terms in Equation 2 represent the costs for $N_{tar}$ ``target'' ($C_{tar}(\theta)$) and $N_{non}$ ``non-target'' speakers ($C_{non}(\theta)$). 

$C_\textrm{llr}$ can be interpreted as a measure that is inversely related to information. The lower the $C_\textrm{llr}$, the more the average information per trial (in bits) increases.  Optimization is performed with the objective of minimizing $C_\textrm{llr}$ loss. $C_\textrm{llr}$ loss is calculated for each minibatch by considering the outputs of the last linear layer as scores and using the class labels to define target and non-target speakers. Thus, $C_\textrm{llr}$ loss minimizes intra-speaker distances by focusing on speaker-specific idiosyncrasies. This is similar to the human approach to ``tell speakers together''. 

\subsubsection{Proposed method: $C_\textrm{llr}$CE loss}
We propose to use the combination of cross-entropy loss and $C_\textrm{llr}$ loss for training ASV systems, so that the loss function can maximize ``inter-speaker'' distances and minimize ``intra-speaker'' distances. We thus use a combined loss function and refer to it as ``$C_\textrm{llr}$CE loss'',

\begin{equation}
    C_\textrm{llr}CE(\theta) = \frac{1}{2} \left( C_\textrm{llr}(\theta) + L_{CE} \right)
\end{equation}

Given that this loss function is inspired by human speaker discrimination strategies in the presence of moderate style variability, i.e, between read and conversational speech, we hypothesize that this loss function will provide the most improvement in conversational speech tasks. 

\section{\label{sec:cllr_experimental_setup}Experimental Setup}
Experiments were setup using Pytorch~\cite{paszke_pytorch_2019} and Kaldi~\cite{povey_kaldi_2011}. Adam~\cite{kingma_adam_2017} optimization was used with a batch size of 128 and trained for 100 epochs.

\subsection{Databases}
\label{ssec:data}
\subsubsection{The UCLA Speaker Variability Database (SVD)}
To systematically study performance in the presence of style variability, the UCLA Speaker Variability Database~\cite{keatingnew2019, kreimanrelationship2015, keating_ldc_2021, keatingucla2021}, a multi-speaker speech database including multiple speech tasks per speaker is employed. It incorporates commonly-occurring variations in speech from 101 female and 101 male speakers, recorded in a sound-attenuated booth at a sampling rate of 22kHz. The tasks include \textbf{reading} sentences characterizing scripted speaking style ($\approx 75$ sec per speaker); giving \textbf{instructions} as unscripted clear monologue style ($\approx 30$ sec  per speaker); \textbf{narrating} a recent happy, annoying, or neutral conversation characterizing unscripted affective speech ($\approx$ 30 sec each affect per speaker); having a conversation on a call with a familiar person (speaker's side speech only) characterizing unscripted \textbf{conversational} style (60--120 sec  per speaker); and talking to pets in a video representing \textbf{pet-directed} speech (60--120 sec  per speaker).

To cover enough phonetic variability, such that there is negligible effects from it, and style variability is predominant~\cite{hasan_duration_2013}, 30~sec-long   speech samples were used for evaluation,  This results in a total of 1,838 30~sec segments for evaluation as shown in Table~\ref{tab:stats}. A majority of speakers had less than 1~min of speech for pet-directed speech and affect-matched narrative case. Hence, the style-matched cases for those styles were omitted, as a style-matched case requires at least 1~min (two 30~sec samples) of speech from the same speaker. This provides a total of 23 style-matched and mismatched tasks for evaluation. The UCLA SVD data  were downsampled to 16~kHz, to match the rest of the databases used.

\subsubsection{The Speakers in the Wild Database (SITW)}
To evaluate the performance of the proposed loss on a large-scale database we use SITW~\cite{mclaren_speakers_2016} for evaluation. It includes speakers employing multiple speaking styles such as interviews, presentation, talk show, social-media videos etc. This database consists of 2,883 recordings from 117 male and 63 female speakers divided into 6,445 utterances sampled at 16~kHz. Single-speaker utterances in the eval set are referred to as ``core''.  Enrollment utterances with multiple speakers (segmentation labels for the person of interest (POI) available) are referred to as ``assist'', while the test utterances that do not include segmentation labels for POI are referred to as ``multi''. 

\subsubsection{VoxCeleb Database} 
 ASV systems were trained on the Voxceleb2 \textit{DEV} set~\cite{chung_voxceleb2_2018}. It consists of speech from YouTube videos of 3,682 male and 2,313 female speakers and includes 1,092,009 utterances with a sampling rate of 16~kHz. The main disadvantage of using VoxCeleb2 for testing is that it comprises interview-style speech only and does not include different styles for each speaker. Hence, we believe that this database does not provide a good representation of the test case scenario targeted in this work.

\section{Results and Discussion \label{sec:results}}
\subsection{UCLA SVD Evaluation \label{sssec:ucla_cllr}}
 The loss functions used in our experiments are cross-entropy loss (CE), $C_\textrm{llr}$ loss, and $C_\textrm{llr}$CE loss. These loss functions are used to train the x-vector system and the best performing VFR conditioning: concatenation with gating. Table~\ref{tab:cllr_loss} compares the performance (in EER \% and minDCF) of the CE and $C_\textrm{llr}$CE loss functions for the UCLA database. The $C_\textrm{llr}$ loss function by itself does not provide an improvement over the widely-used CE loss function in both the x-vector and VFR conditioning architectures.  Therefore, we do not report those results in Table 2. Statistical significance was verified using McNemar's test~\cite{mcnemar_note_1947}. Unless mentioned explicitly, all performance differences reported in this section are significant with $p<0.05$.
 

In the x-vector setup, $C_\textrm{llr}$CE loss provides statistically significant improvements over CE loss in 11/23 tasks and is the same as CE loss in 12/23 tasks. The minDCF is significantly better with $C_\textrm{llr}$CE loss in 7/23 tasks and worse in 4/23 tasks when compared to CE loss. 
In the VFR conditioning setup, $C_\textrm{llr}$CE loss provides  statistically significant improvements in 4/23 tasks, especially in tasks involving conversational style, compared to VFR conditioning with CE loss. Since in the VFR conditioning setup style variability is addressed, the improvement with $C_\textrm{llr}$CE loss is not consistent. The CE loss performs significantly better in 8/23 tasks. The performance in terms of minDCF values show that the $C_\textrm{llr}$CE loss provides significant improvements over CE loss in 12/23 tasks, and the same performance in 11/23 tasks. The most relative improvement is seen with tasks that include conversational or narrative style speech in enrollment and/or test conditions. 

Overall, VFR conditioning trained with $C_\textrm{llr}$CE loss provides significant improvements over the x-vector baseline (with CE loss) in 7/23 tasks in EER, and 12/23 tasks in minDCF. When compared to their CE counterparts, we again notice that the conditions where the $C_\textrm{llr}$CE loss provides improvements are the ones that include conversation style speech and narrative style speech (closest to the conversation style). 


\subsection{SITW Evaluation}

Table~\ref{tab:sitw_cllr} presents the performance on the SITW evaluation set using different loss functions. The loss functions used are cross-entropy loss (CE), $C_\textrm{llr}$ loss, and $C_\textrm{llr}$CE loss. These loss functions are used to train the x-vector system and the best performing VFR conditioning: concatenation with gating. 

The results show that the best performing system in terms of minDCF values is the one with combination loss in the VFR conditioning setup. However, EER values of the $C_\textrm{llr}$CE loss in the VFR conditioning setup are slightly worse than the CE loss counterpart for assist-core and assist-multi evaluations. Overall, the proposed loss function with the VFR conditioning setup results in the best performance on the SITW evaluation. Since SITW involves mainly conversational speech, this result agrees with our hypothesis that the new loss function improves ASV system performance for conversational styles. 

Overall results show that the combined $C_\textrm{llr}$CE loss improves ASV performance for the two configurations when compared to CE and $C_\textrm{llr}$ loss functions individually. Thus, implying that the $C_\textrm{llr}$ and CE loss functions are complementary. 

\section{Conclusion\label{sec:conclusion}}
In order to improve ASV performance in the presence of style variability, this work introduces a new loss function ($C_\textrm{llr}$CE) that is inspired by human speech perception. $C_\textrm{llr}$CE loss focuses on both speaker-specific idiosyncrasies to ``tell speakers together'' and on relative acoustic distances between the speakers to ``tell speakers apart''. This combined loss maximizes inter-speaker distances while minimizing intra-speaker distances resulting in performance improvements over the widely used CE loss function and also the $C_\textrm{llr}$ loss function, showing their complementarity.  To the best of our knowledge, this is the first work to propose a training loss function for ASV that is inspired by human perception. In future, this work will be extended to study perception strategies between other styles and use those to improve ASV approaches. Further studies on the effects on short-duration scenario~\cite{guo_speaker_2016, ravi_exploring_2020} and other embedding extractors~\cite{zhou_resnext_2021, desplanques_ecapa-tdnn_2020} would provide better understanding of the proposed loss function. 

\bibliographystyle{IEEEtran}

\bibliography{refs}


\end{document}